\documentclass[aps,twocolumn,amsmath,amssymb,showpacs,prb,superscriptaddress,unsortedaddress]{revtex4}
\usepackage{epsf}
\usepackage{graphicx}

\begin{document}

\title{Origins of large critical temperature variations in single layer cuprates}

\author{A.~D.~Palczewski}
\affiliation{Ames Laboratory and Department of Physics and Astronomy, Iowa State University, Ames, IA 50011, USA}

\author{T.~Kondo}
\affiliation{Ames Laboratory and Department of Physics and Astronomy, Iowa State University, Ames, IA 50011, USA}

\author{R.~Khasanov}
\affiliation{Laboratory for Muon Spin Spectroscopy, Paul Scherrer
Institut, CH-5232 Villigen PSI, Switzerland}

\author{N.~N.~Kolesnikov}
\affiliation{Institute of Solid State Physics, Chernogolovka, 142432 Russia}

\author{A.~V.~Timonina}
\affiliation{Institute of Solid State Physics, Chernogolovka, 142432 Russia}

\author{E.~Rotenberg}
\affiliation{Advanced Light Source, Berkeley National Laboratory, Berkeley, California 94720, USA}

\author{T.~Ohta}
\affiliation{Advanced Light Source, Berkeley National Laboratory, Berkeley, California 94720, USA}

\author{A.~Bendounan}
\affiliation{Labratory for Neutron Scattering, ETH Zurich and Paul Scherrer Institute, CH-5232 Villigen PSI, Switzerland}

\author{Y.~Sassa}
\affiliation{Labratory for Neutron Scattering, ETH Zurich and Paul Scherrer Institute, CH-5232 Villigen PSI, Switzerland}

\author{A.~V.~Fedorov}
\affiliation{Advanced Light Source, Berkeley National Laboratory, Berkeley, California 94720, USA}

\author{S.~Pailh\'es}
\affiliation{Labratory for Neutron Scattering, ETH Zurich and Paul Scherrer Institute, CH-5232 Villigen PSI, Switzerland}

\author{A.~F.~Santander-Syro}
\address{Laboratoire Photons Et Mati\`ere, UPR-5 CNRS, ESPCI, 10 rue Vauquelin, 75231 Paris cedex 5, France}
\address{Labratoire de Physique des Solides, UMR-8502 CNRS, Universit\'e Paris-Sud, B\^at. 510, 
                 91405 Orsay, France}

\author{J.~Chang}
\affiliation{Labratory for Neutron Scattering, ETH Zurich and Paul Scherrer Institute, CH-5232 Villigen PSI, Switzerland}

\author{M.~Shi}
\affiliation{Swiss Light Source, Paul Scherrer Institute, CH-5232 Villigen PSI, Switzerland}

\author{J.~Mesot}
\affiliation{Labratory for Neutron Scattering, ETH Zurich and Paul Scherrer Institute, CH-5232 Villigen PSI, Switzerland}

\author{H.~M.~Fretwell}
\affiliation{Ames Laboratory and Department of Physics and Astronomy, Iowa State University, Ames, IA 50011, USA}

\author{A.~Kaminski}
\affiliation{Ames Laboratory and Department of Physics and Astronomy, Iowa State University, Ames, IA 50011, USA}

\date{\today}
\begin{abstract}

We study the electronic structures of two single layer superconducting cuprates, Tl$_2$Ba$_2$CuO$_{6+\delta}$ (Tl2201) and (Bi$_{1.35}$Pb$_{0.85}$)(Sr$_{1.47}$La$_{0.38}$)CuO$_{6+\delta}$ (Bi2201) which have very different maximum critical temperatures (90K and 35K respectively) using Angular Resolved Photoemission Spectroscopy (ARPES). We are able to identify two main differences in their electronic properties.  First, the shadow band that is present in double layer and low T$_{c,max}$ single layer cuprates is absent in Tl2201. Recent studies have linked the shadow band to structural distortions in the lattice and the absence of these in Tl2201 may be a contributing factor in its T$_{c,max}$.
Second, Tl2201's Fermi surface (FS) contains long straight parallel regions near the anti-node, while in Bi2201 the anti-nodal region is much more rounded.  Since the size of the superconducting gap is largest in the anti-nodal region, differences in the band dispersion at the anti-node may play a significant role in the pairing and therefore affect the maximum transition temperature.

\end{abstract}


\maketitle

Despite more than 20 years of effort, there is still no consensus on what is the nature of the superconducting coupling mechanism in the high T$_c$ superconductors. Early theoretical works\cite{Anderson 1997} proposed that interlayer interactions between the copper oxygen (Cu-O) planes in these quasi 2D materials played a key role in the pairing mechanism.  However, some predictions from this model were later found to be inconsistent with experiment\cite{Tsvetkov 1998}.
Yet, there remains empirical evidence that both the maximum transition temperature (T$_{c,max}$) and the size of the superconducting gap of the high temperature superconducting cuprates (HTSC) depends, sometimes strongly, on the number of Cu-O layers per unit cell\cite{T. Sato 2002}.  Bismuth\cite{D. L. Feng 2002}, thallium\cite{Y. C. Ma 2005}, and mercury\cite{C. Ambrosch-Draxl 2004} -based cuprates all show an increase in T$_{c,max}$ with the number of Cu-O layers.
While  T$_{c,max}$ increases with the number of Cu-O layers (peaking at 3 layers per unit cell), it is not always the same for a given number of layers. In particular, there are two single layer materials, Tl$_2$Ba$_2$CuO$_{6+\delta}$(Tl2201)\cite{C. C. Torardi 1988} and HgBa$_2$CuO$_{4+\delta}$(Hg1201)\cite{W.S. Lee 2006} (T$_{c,max}$$\sim$90K), whose transition temperatures are actually closer to that of other double layer cuprates.
This could mean that either T$_{c,max}$ is somehow enhanced in Tl2201 and Hg1201 or that T$_{c,max}$ for all single layer cuprates is intrinsically closer to 95K and other mechanisms, for example, lattice distortions in the Bismuth-based materials\cite{K. Nakayama 2006} reduce T$_{c,max}$. One can imagine that adding more Cu-O layers per unit cell to the Bi-based material (going from Bi2201 to Bi2212) creates an additional channel, thereby enhancing the superconductivity and pushing the T$_{c,max}$ back up to $\sim$ 90K. 
To help explore these ideas and explain the large variation of T$_{c,max}$ of the single layer compounds, it is essential to look for differences in their electronic structure through Angular-Resolved Photoemission Spectroscopy (ARPES)\cite{A. Damascelli 2003,J. C. Campuzano 2004}.

%%%%%%%%%%%%%%%%%%%%%Crystal structure
\begin{figure}
\includegraphics[width=3.6in]{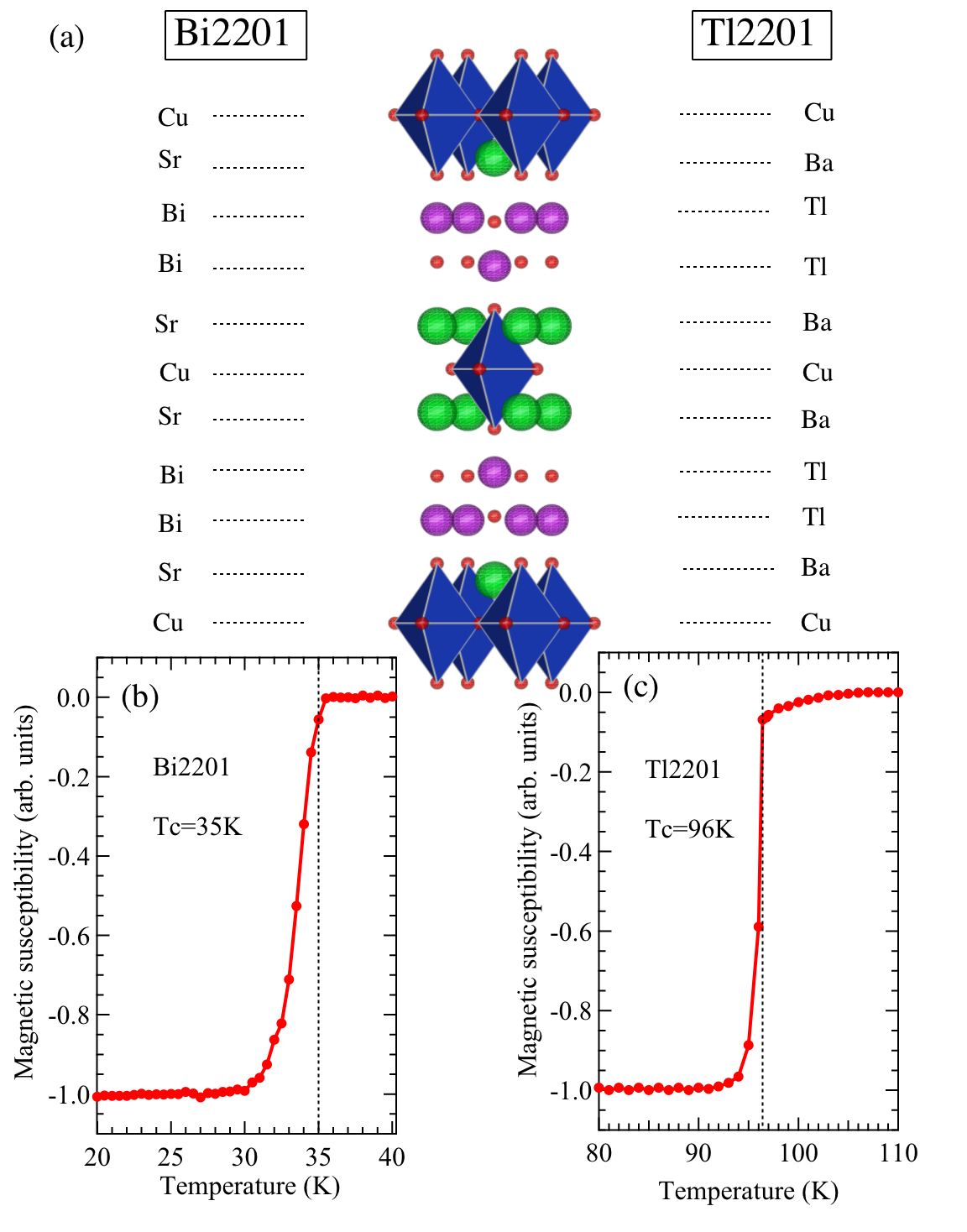}
\caption{(Color Online) (a) Schematic structure of Bi$_2$Sr$_2$CuO$_{6+\delta}$(Bi2201) and Tl$_2$Ba$_2$CuO$_{6+\delta}$(Tl2201). The small red dots represent oxygen atoms and the other larger atoms are labeled by the symbol on the left (Bi2201) or right (Tl2201).  Each layer is made up of the particular atoms bounded to oxygen, with the double pyramids representing copper oxygen bonds. (b)-(c) SQUID magnetization curves for Bi2201 and Tl2201.}
\label{Fig. 1}
\end{figure}
%%%%%%%%%%%%%%%%%%%%%%

Here we report an ARPES study on the electronic structure of two single layer cuprates with distinctly different maximum critical temperatures: Tl$_2$Ba$_2$CuO$_{6+\delta}$ (Tl2201) T$_c$$\sim$90K and (Bi$_{1.35}$Pb$_{0.85}$)(Sr$_{1.47}$La$_{0.38}$)CuO$_{6+\delta}$ (Bi2201) T$_c$$\sim$35K.
We find two striking differences in the Fermi surface (FS) maps at the chemical potential. First, the shadow band (usually attributed to structural distortions\cite{K. Nakayama 2006,A. Mans 2006,A. Koitzsch 2004}) is present in single layer Bi2201 (and double layer Bi2212) but is absent in Tl2201.
Second (and possibly more important), the FS of Tl2201 has long parallel ``nested" regions close to the antinodes (where the superconducting gap reaches its maximum value). This feature is very similar to that found in double layered Bi2212 with a T$_{c,max} of $$\sim$ 90K, while it is absent in Bi2201.
In other words, materials with a high T$_c$ have strongly nested FS.

%%%%%%%%%%%%%%%%%%%%%FS picture
\begin{figure}
\includegraphics[width=3.6in]{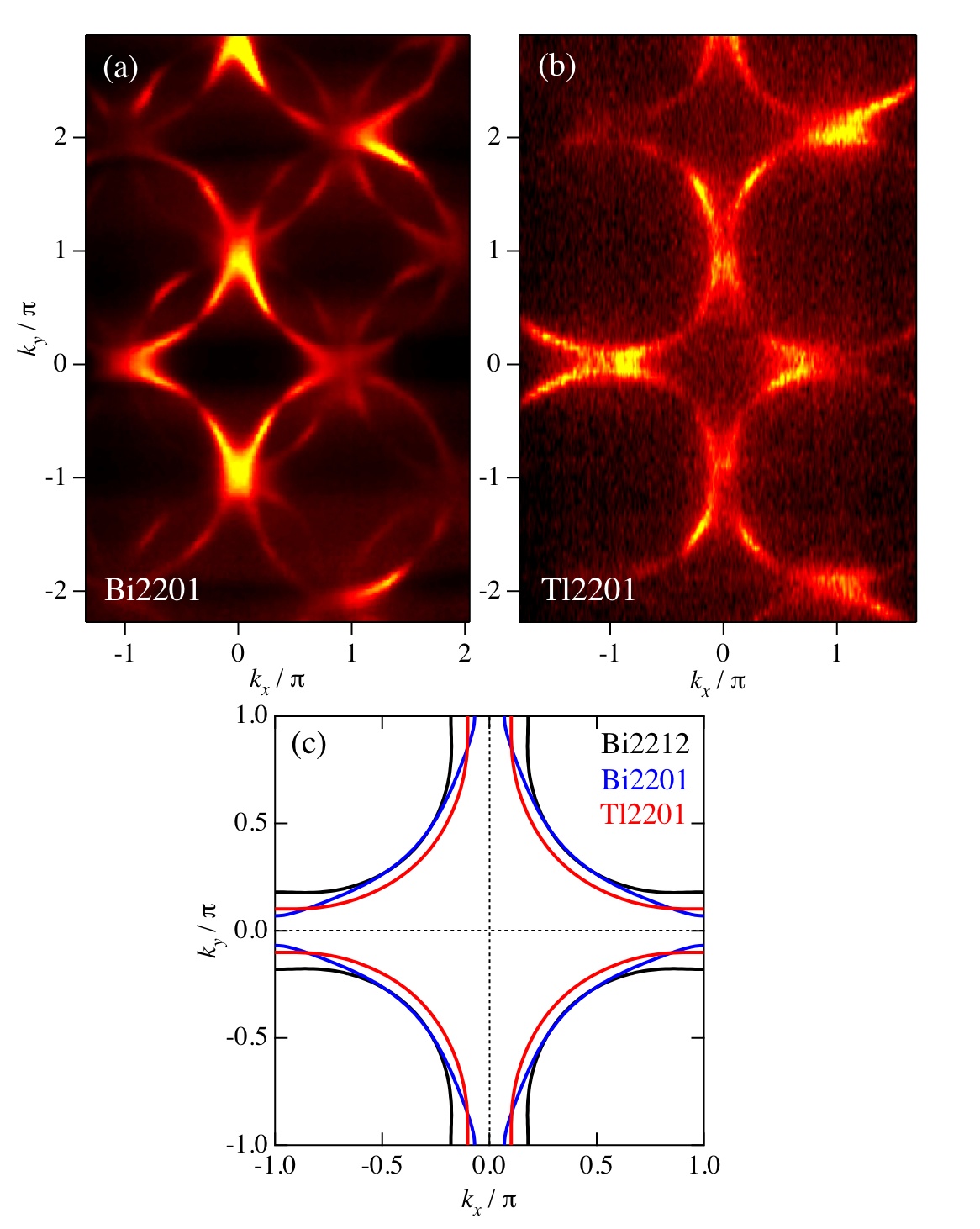}
\caption{(Color Online) Intensity at the Fermi energy in multiple Brillouin zones for (a) Bi2201 and (b) Tl2201. All data was collected at a photon energy of 105 eV. High (low) intensity regions appear bright (dark) in the color map. (c) Tight binding fitting plots,  Bi2212\protect\cite{M. R. Norman 1995} (black), Bi2201 (blue) and Tl2201 (red), fitting parameters for (c) are found in Table 1.}
\label{Fig. 2}
\end{figure}
%%%%%%%%%%%%%%%%%%%%%%

Optimally doped Bi2201 single crystals were grown using the floating zone (FZ) method\cite{T. Kondo 2005}.  The substitution of Pb suppresses the modulation in the Bi-O layers\cite{H. Ding 1996} that normally causes complications (superlattice) in interpreting the band structure in pristine Bi$_2$Sr$_2$CuO$_{6+\delta}$ \cite{A. Damascelli 2003, J. C. Campuzano 2004}.  Near optimally doped Tl$_2$Ba$_2$CuO$_{6+\delta}$ crystals were grown in an air atmosphere inside zirconium dioxide multilayered crucibles\cite{N.N. Kolesnikov 1992,N.N. Kolesnikov 1995}. Single crystals samples of both materials used in ARPES experiments are of exceptional quality as evidenced by very sharp superconducting transitions with typical  widths $\sim$2-4K shown in Fig. 1(b)-(c).
FS measurements for Tl2201 were performed at the Swiss Light Source (SLS) on beamline X09LA-HRPES with a Scienta SES2002 at 49 eV photon energy. The choice of photon energy was dictated by need to maximize both the signal intensity and energy resolution. As evident from Fig. 5 there are two main energy for which the signal reaches maximum: 49 eV and 74 eV.  The signal is certainly stronger when using the latter, however due to characteristics of beamline the energy resolution there would be significantly reduced. The energy and angular resolution was set to 30 meV and 0.5$^\circ$ respectively.
Electronic structure information for Bi2201 and Tl2201 was acquired at the Advanced Light source (ALS) on Beamline 7.0.1 with the SCIENTA R4000 analyzer at 105 eV photon energy. The energy and angular resolution of the R4000 was set to 40 meV and 0.5$^\circ$ respectively. 
Tl2201 photon energy dependence data was taken at the ALS on beamline 12.0.1.1 using a SCIENTA 100 analyzer.  The energy and angular resolutions was set to 50 meV and 0.3$^\circ$ respectively.
Bi2201 doping dependence data was acquired on a Scienta SES2002 hemispherical analyzer using a Gammadata VUV5000 photon source (HeI$\alpha$) at Iowa State University. The energy and angular resolution was set to 5meV and 0.13$^\circ$ respectively. 
All data was acquired on \textit{in situ} cleaved crystals at or below 20K under UHV, with the samples being kept at their cleaving temperature throughout the measurement process.  During  the measurement process we had to cleave multiple Tl2201 samples in order to get reliable and reproducible results.  This was mainly due to Tl2201's inability to cleave nicely.  Bi2201 on the other hand almost always cleaves nicely, so multiple cleaves were not as important.

%%%%%%%%%%%%%%%%%%%%%%%%%%tight binding parameters table
\begin{table*} [htdp]
\caption{Tight-Binding fitting function ($\varepsilon (\vec k)$) and experimental fit for Bi2201, Tl2201 and Bi2212\protect\cite{M. R. Norman 1995} 
where $\varepsilon (\vec k) = \sum {c_i \eta _i (\vec k)} $}
\begin{tabular}{||c|ccc||}
\hline
 $\eta _i (\vec k)$ & $c_i$ $\textrm{Bi2201}$  &  $c_i$ $\textrm{Tl2201}$   & $c_i$ $\textrm{Bi2212}$\\
\hline
1& 0.16895 $\pm$ 0.013 &  0.24103 $\pm$ 0.0202 & 0.1305\\
$\frac{1}{2}$($\cos$ \textit{\textit{k$_x$}}+$\cos$ \textit{k$_y$}) & -0.73338 $\pm$ 0.0161 & -0.72153 $\pm$ 0.0328 & -0.5951\\
$\cos$ \textit{k$_x$}$\times$$\cos$ \textit{k$_y$}   & 0.11389 $\pm$ 0.00786 & 0.14813 $\pm$ 0.00935 & 0.1636\\
$\frac{1}{2}$($\cos$ 2\textit{k$_x$}+$\cos$ 2\textit{k$_y$})    & -0.11086 $\pm$ 0.00573  & -0.17287 $\pm$ 0.0115 & -0.0519\\
$\frac{1}{2}$($\cos$ 2\textit{k$_x$}$\times$$\cos$ 2\textit{k$_y$}+$\cos$ \textit{k$_x$}$\times$$\cos$ 2\textit{k$_y$}) & -0.049688 $\pm$ 0.0248 & -0.01604 $\pm$ 0.0359 & -0.1117\\
$\cos$ 2\textit{k$_x$}$\times$$\cos$ 2\textit{k$_y$}   & 0.045032 $\pm$ 0.00751& 0.048246 $\pm$ 0.016 & 0.051\\ [1ex]
\hline
\end{tabular}
\label{Table 1}
\end{table*}
%%%%%%%%%%%%%%%%%%%%%%%%%%%%%%%%

%%%%%%%%%%%%%%%%%%%%%
\begin{figure}
\includegraphics[width=3.6in]{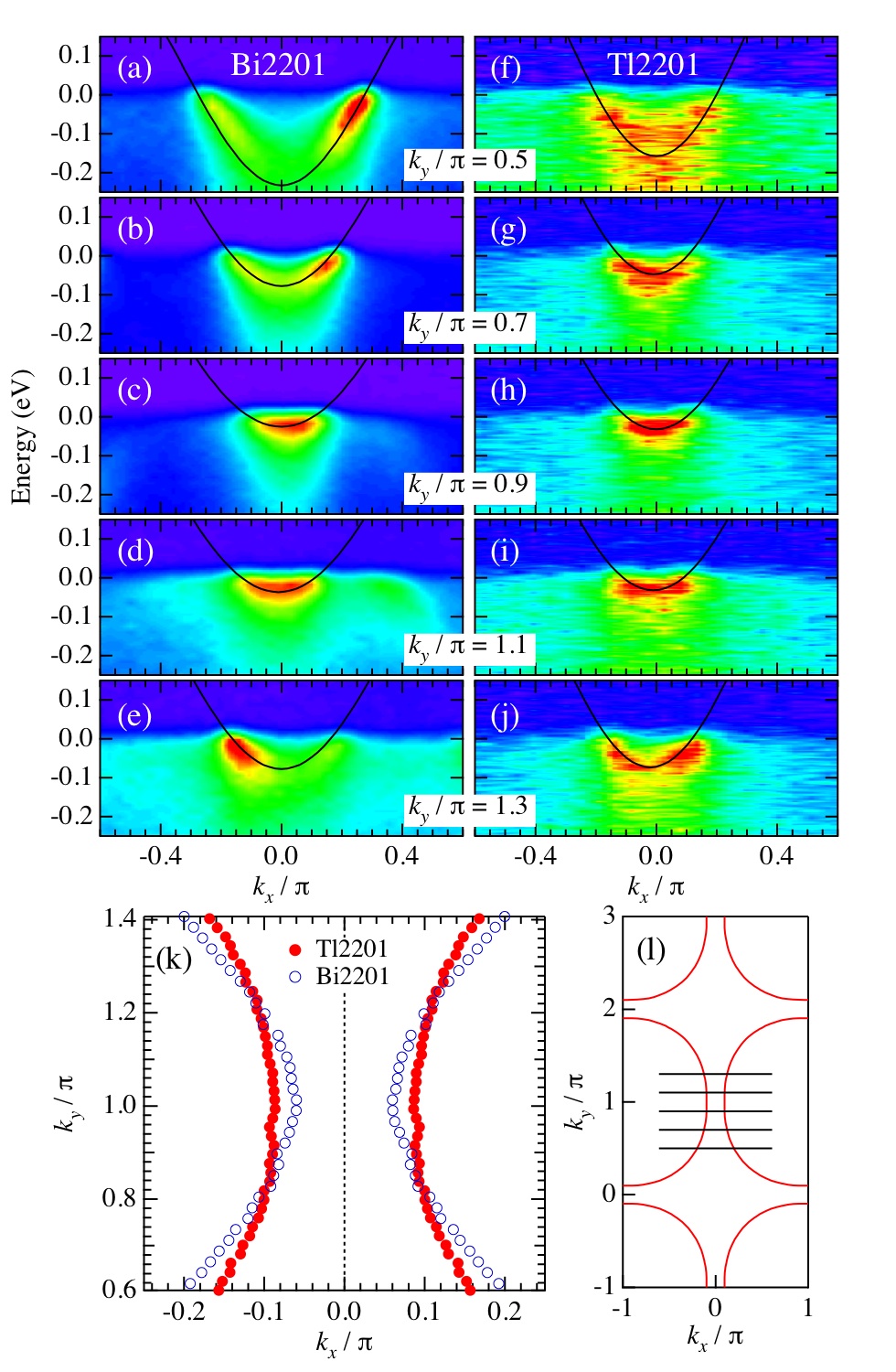}
\caption{(Color Online) Momentum distribution curve for (a)-(e) Bi2201 and (f)-(j) Tl2201 taken at k$_y$/$\pi$=0.5,0 .7,0 .9, 1.1 and 1.3. The lowest intensity corresponds to red while the highest intensity corresponds to dark blue moving through the color spectrum. The colored pictures are the original APRES data while the black lines are tight-binding fit.  The tight binding fitting parameters for the black lines are located in Table 1, (k) FS taken from peak position of MDC for Bi2201 (blue dots) and Tl2201 (red crosses), (l) schematic MDC location for (a)-(j).}
\label{Fig. 3}
\end{figure}
%%%%%%%%%%%%%%%%%%%%%%

The schematic crystal structure of Tl2201 and Bi2201 are shown in Fig. 1a\cite{Shaked 1994}.  Each material's unit cell contains a single Cu-O layer with dual layers of Tl-O and Ba-O (Tl2201) or Bi-O and Sr-O (Bi2201).   We note that Tl2201 has a tetragonal (i.e.\textit{ a=b}) structure with nearly perfectly flat Cu-O layers and a slight buckling in the Tl-O and Ba-O layers\cite{N.N. Kolesnikov 1995}.  In contrast, Bi2201's structure has a degree of orthorombicity (i.e. \textit{a$\cong$b}) accompanied by buckling in all layers\cite{C. C. Torardi 1988}.  The two materials also have very different cleaving properties. Tl2201 has strong bonding between the layers, which makes it difficult to cleave, often leaving behind a rather rough surface.  Whereas, Bi2201 is very well known for excellent cleaving properties and is material of choice for surface studied such as ARPES or STM/STS. This is because the bonding between adjacent Bi-O layers is due to Van der Waals interaction. In majority of cases after cleaving we were able to obtain a flat mirror-like surfaces.
%%%%%%%%%%%%%%%%%%%%%
\begin{figure}
\includegraphics[width=3.6in]{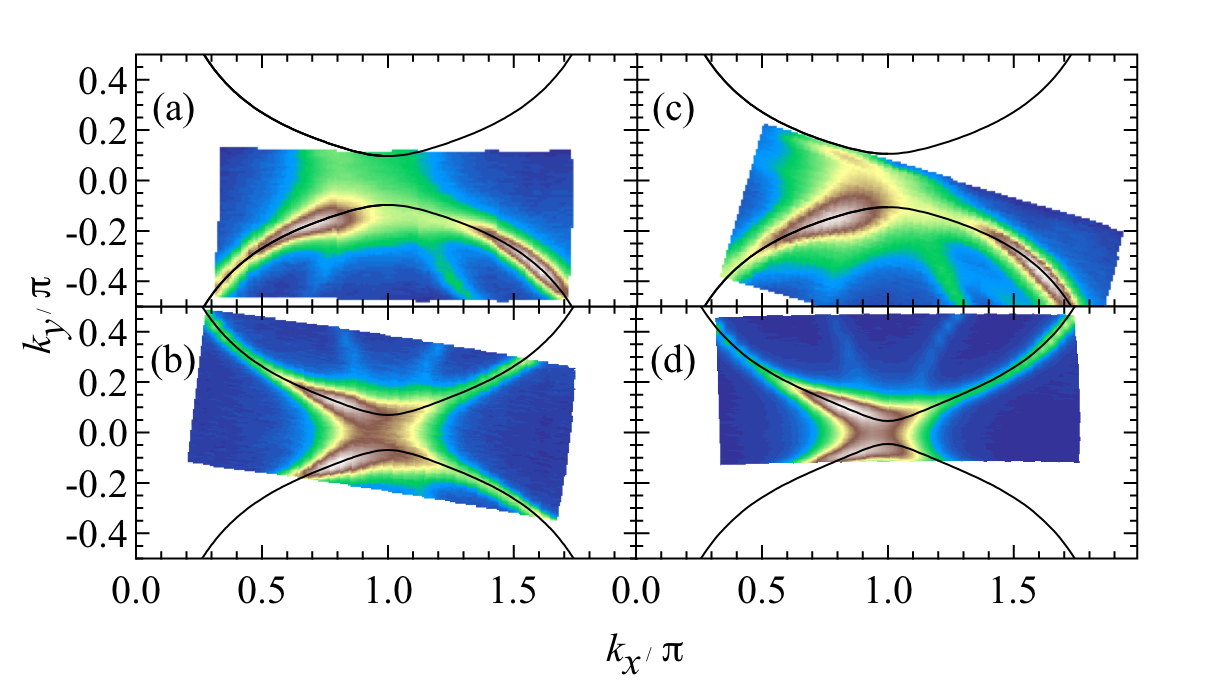}
\caption{(Color Online) Intensity maps of Bi2201 taken around  ($\pi$,0) for different carrier concentrations (a) 0.23, (b) 0.25, (c) 0.27, and (d) 0.29, with black line represents the tight binding fits for each doping level}
\label{Fig. 4}
\end{figure}
%%%%%%%%%%%%%%%%%%%%%%

The ARPES intensity integrated from 20meV to -40meV about the chemical potential is plotted as a function of momentum for Bi2201 and Tl2201 in Fig. 2 (a) and (b) respectively. The bright areas correspond to high intensity and represent the Fermi surface (FS) - those locations in momentum space where the band crosses the chemical potential.  One can see that both FS are similar to the usual calculations of a Cu-O layer inside a cuprate\cite{H. Krakauer 1988,O. K. Anderson 1995}, with a couple of distinct differences.
First, the shadow band, found in some cuprates\cite{K. Nakayama 2006,S. V. Borisenko 2000,A. Koitzsch 2004} including single layer Bi2201 (T$_{c,max}$=35K, left panel) and LSCO (T$_{c,max}$=40K) as well as two layer Bi2212 (T$_{c,max}$$\cong$90K), is absent in single layer Tl2201 (T$_{c,max}$$\cong$90K, right panel).  
The second more subtle difference is the shape of the FS close to the antinode ($\pi$,0). To better compare the shape of the FS, we have performed a tight binding analysis on each of our samples, the results from these fits are shown in Fig. 2 (c). The fitting analysis was performed using full 3D band dispersion data, examples of which are shown in Fig. 3. We also present the published tight-biding fits for Bi2212\cite{M. R. Norman 1995} in Fig. 2 (c) for comparison. Fitting parameters for all three cases are presented in Table 1.  Based on these parameters we have calculated the carrier concentration level for the three systems:  0.17 for Bi2212, 0.27 for Bi2201 and 0.35 for Tl2201. The shape of the FS for Tl2201 and Bi2212 are almost identical; the only visual difference between the two arises from the differences in their carrier concentrations.  They both display long, nearly parallel FS segments close to the antinode. The FS of Bi2201 is quite different in this region of momentum space.  Bi2201 FS is much more rounded with no significant parallel segments.  We have to point out that the length of the parallel segments in the anti-nodal regions will, in principle, depend on carrier concentration. In heavily overdoped cuprates, the antinodal regime of the FS can become less parallel and eventually close (disappearing completely from the FS)\cite{A. Kaminski 2006}.  In our case, Tl2201 has a higher carrier concentration (more overdoped) than the Bi2201, yet Tl2201's antinodal FS nesting is still much greater than in Bi2201.  To show that Bi2201's rounded FS is not a doping dependent feature but a fundamental characteristic, we present Fig. 4.  Moving from top to bottom and left to right, i.e. (a)-(d), we show the FS of Bi2201 about ($\pi$,0) at carrier concentration levels of 0.23, 0.25, 0.27, 0.29 respectively.  We see the shape changes slightly as we change doping, as is expected yet, the general roundness remained throughout all dopings levels.

Fig. 5 (a) shows the peak intensity verse photon energy for Tl2201 taken at a constant region of momentum space near ($\pi$,0). The variation in the intensity arises from the matrix element effect\cite{Bansil 1999} during the photoemission process.  Fig. 5 (b)-(d) shows how the matrix elements can affect the overall dispersion with some energies being better than other for data acquisition.  From this curve we have identified 49 eV and 74 eV as the best energies for obtaining high resolution data from Tl2201 samples.

We now discuss why T$_{c,max}$ is much higher in Tl2201 compared to Bi2201.
First, Tl2201 has a tetragonal crystal structure with flat Cu-O layers, whereas Bi2201 is orthorhombic with buckled Cu-O layers\cite{C. C. Torardi 1988}. It is known that local lattice distortions (produced by chemical inhomogeneity) can reduce the value of T$_c$ in a systematic way\cite{H. Eisaki 2004}. It has also been shown that the larger the Cu-O plane buckling angle, the lower the T$_{c,max}$\cite{S Kambe 2001}. So, distortions in the Cu-O planes have long been known to cause a lowering of the T$_{c,max}$.
Second, our results show that the FS of Tl2201 does not contain a shadow band, but Bi2201 does. There are several explanations for the origins of the shadow band. The most convincing explanation to date is that it is due to structural distortions, either in the form of an orthorhombic distortion of the lattice\cite{K. Nakayama 2006}, and/or by diffraction of the outgoing photoelectron by the superstructure of the BiO layer at the surface. Whatever the cause, if the shadow band is absent (as in Tl2201), it suggests that the material is free of the structural distortions that could potentially lower the T$_{c,max}$
Finally, Tl2201 has strong interlayer interactions that are absent in Bi2201. The same strong interlayer bonding is also present in another high T$_{c,max}$ single layer cuprate Hg1201 (T$_{c,max}$$\sim$95K)\cite{W.S. Lee 2006}.
Given the above, our observation that Tl2201 does not exhibit a shadow band is fully consistent with the absence of structural distortions of its lattice and its unusually high T$_{c,max}$.
We now address the fact that Bi2212 is known to have buckled Cu-O planes, orthorhombic distortions, a shadow band and weak interlayer interactions, yet it still has a high T$_{c,max}$, which is comparable to that of Tl2201.
We speculate that the extra Cu-O layer per unit cell in Bi2212 enhances the superconductivity and raises the T$_{c,max}$.  This has been seen in other multi-layered cuprates where Cooper pairs are allowed to tunnel between the Cu-O layers through Josephson coupling, raising T$_{c,max}$\cite{S. Chakravarty 2004,T. A. Zaleski 2005}.

%%%%%%%%%%%%%%%%%%%%%photon energy dependence
\begin{figure}
\includegraphics[width=3.2in]{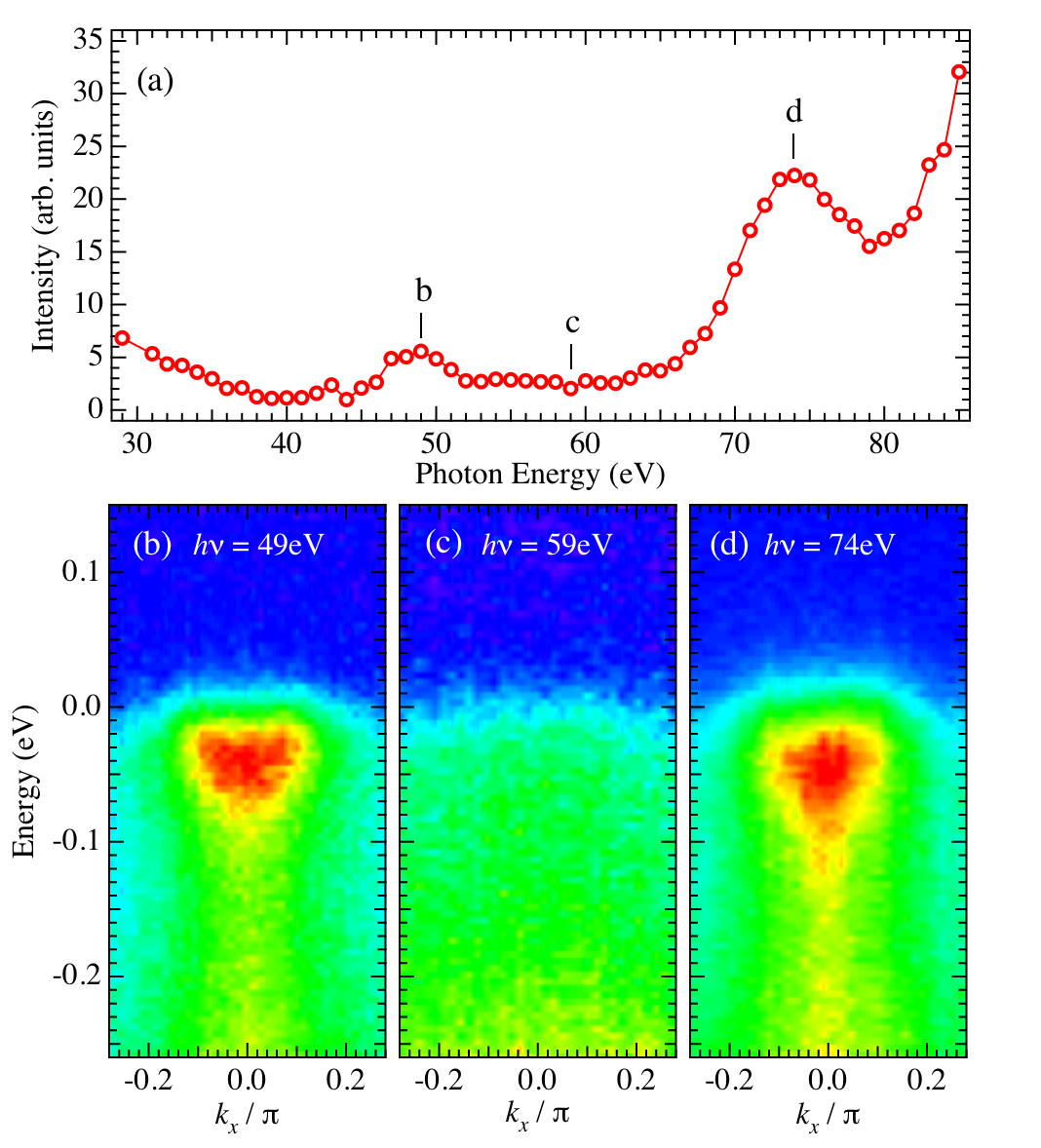}
\caption{(Color Online) (a) Photon energy dependence for Tl2201 taken at around ($\pi$,0).  Markers in (a) are at 49eV, 59eV and 74eV respectively, corresponding to the energy momentum cuts in (b), (c), and (d).  In (b)-(d) dark blue corresponds to low intensity, while red corresponds to high intensity moving through the color spectrum}
\label{Fig. 5}
\end{figure}
%%%%%%%%%%%%%%%%%%%%%%

Finally, our data shows a relationship between the length of the long parallel (nested) FS segments centered about ($\pi$,0) and T$_{c,max}$.  Looking back to Fig. 2 (c) we see that Tl2201 and Bi2212 have very similar nested FS segments and approximately the same T$_{c,max}$. In contrast, Bi2201's FS segments are much rounder with a lower T$_{c,max}$.   Our data suggests that FS nesting at the antinode is related to the enhanced T$_{c,max}$.  We also note that the superconducting and pseudo-gaps reach a maximum in these regions\cite{J. Mesot 1999,H. Ding 1995}, with other studies suggesting this region is critical in understanding of how cuprate superconductivity works\cite{Kondo 2007,Hanaguri 2004,Shen 2005,Hanaguri 2007}.  Our observation of significant FS nesting in Tl2201 is an important new result.

In conclusion, we report a comparative study on the electronic structures of two single layer cuprates Tl2201 T$_{c,max}$$\sim$90K and Bi2201 T$_{c,max}$$\sim$35K, along with photon energy data for Tl2201.  We find two striking differences in the occurrence of the shadow band and the shape of the FS close to the antinodes.  First, the shadow band in single layer Bi2201 and double layer Bi2212 is absent in Tl2201.  Second Tl2201 has long parallel (nested) regions on its FS (similar to double layer Bi2212 with T$_{c,max}$$\sim$90K), while these regions are much smaller (if not absent) in low T$_{c,max}$ Bi2201. Our data shows two non trivial results for superconducting cuprates.  First, there may be a balance between structural distortions and interlayer interactions that help control T$_{c,max}$ in the cuprates.  Second, there is a qualitative relationship between the length of the anti-nodal nesting and T$_{c,max}$ in our cuprates.

The work at the Ames Laboratory was supported by the Department of Energy at Iowa State University.  Ames laboratory is supported under Contract No. DE-AC02-07CH11358. The Advanced Light Source is supported by the Director, Office of Science, Office of Basic Energy Sciences, of the U.S. Department of Energy under Contract No. DE-AC02-05CH11231. This research project is also supported by the European Commission under the 6th Framework Programme: Strengthening the European Research Area, Research Infrastructures. Contract n°: RII3-CT-2004-506008'.

\end{document}